\documentclass[11pt,a4paper]{article}
\pdfoutput=1
\usepackage{amssymb} \usepackage{amsmath} \usepackage{graphicx}
\usepackage{epsfig,latexsym,color,xcolor}
\usepackage{ulem}
\usepackage{amstext}    
\usepackage{array} 
\usepackage{slashed}
\begin{document}

\def\Giulia{\bf\color{red}}
\def\bef{\begin{figure}}
\def\eef{\end{figure}}
\newcommand{\ans}{ansatz }
\newcommand{\be}[1]{\begin{equation}\label{#1}}
\newcommand{\beq}{\begin{equation}}
\newcommand{\ee}{\end{equation}}
\newcommand{\beqn}[1]{\begin{eqnarray}\label{#1}}
\newcommand{\eeqn}{\end{eqnarray}}
\newcommand{\bd}{\begin{displaymath}}
\newcommand{\ed}{\end{displaymath}}
\newcommand{\mat}[4]{\left(\begin{array}{cc}{#1}&{#2}\\{#3}&{#4}
\end{array}\right)}
\newcommand{\matr}[9]{\left(\begin{array}{ccc}{#1}&{#2}&{#3}\\
{#4}&{#5}&{#6}\\{#7}&{#8}&{#9}\end{array}\right)}

\newcommand{\matrr}[6]{\left(\begin{array}{cc}{#1}&{#2}\\
{#3}&{#4}\\{#5}&{#6}\end{array}\right)}
\newcommand{\cvb}[3]{#1^{#2}_{#3}}
\def\lsim{\raise0.3ex\hbox{$\;<$\kern-0.75em\raise-1.1ex
e\hbox{$\sim\;$}}}
\def\gsim{\raise0.3ex\hbox{$\;>$\kern-0.75em\raise-1.1ex
\hbox{$\sim\;$}}}
\def\abs#1{\left| #1\right|}
\def\simlt{\mathrel{\lower2.5pt\vbox{\lineskip=0pt\baselineskip=0pt
           \hbox{$<$}\hbox{$\sim$}}}}
\def\simgt{\mathrel{\lower2.5pt\vbox{\lineskip=0pt\baselineskip=0pt
           \hbox{$>$}\hbox{$\sim$}}}}
\def\unity{{\hbox{1\kern-.8mm l}}}
\newcommand{\eps}{\varepsilon}
\def\ep{\epsilon}
\def\ga{\gamma}
\def\Ga{\Gamma}
\def\om{\omega}
\def\omp{{\omega^\prime}}
\def\Om{\Omega}
\def\la{\lambda}
\def\La{\Lambda}
\def\al{\alpha}
\def\beq{\begin{equation}}
\def\eeq{\end{equation}}
\newcommand{\MyRed}{\color [rgb]{0.8,0,0}}
\newcommand{\MyGreen}{\color [rgb]{0,0.7,0}}
\newcommand{\MyBlue}{\color [rgb]{0,0,0.8}}
\def\GV#1{{\MyRed [GV: #1]}}
\def\MB#1{{\MyGreen [MB: #1]}}   
\def\GAV#1{{\MyBlue [GAV: #1]}}  
\newcommand{\sect}[1]{\setcounter{equation}{0}\section{#1}}
\renewcommand{\theequation}{\thesection.\arabic{equation}}
\newcommand{\ov}{\overline}
\renewcommand{\to}{\rightarrow}
\renewcommand{\vec}[1]{\mathbf{#1}}
\newcommand{\vect}[1]{\mbox{\boldmath$#1$}}
\def\tm{{\widetilde{m}}}
\def\mcirc{{\stackrel{o}{m}}}
\newcommand{\Dm}{\Delta m}
\newcommand{\dm}{\varepsilon}
\newcommand{\tanb}{\tan\beta}
\newcommand{\nbar}{\tilde{n}}
\newcommand\PM[1]{\begin{pmatrix}#1\end{pmatrix}}
\newcommand{\up}{\uparrow}
\newcommand{\down}{\downarrow}
\newcommand{\refs}[2]{eqs.~(\ref{#1})-(\ref{#2})}
\def\omE{\omega_{\rm Ter}}
\newcommand{\eqn}[1]{eq.~(\ref{#1})}
%

\newcommand{\DSUSY}{{SUSY \hspace{-9.4pt} \slash}\;}
\newcommand{\DCP}{{CP \hspace{-7.4pt} \slash}\;}
\newcommand{\mc}{\mathcal}
\newcommand{\gr}{\mathbf}
\renewcommand{\to}{\rightarrow}
\newcommand{\gtc}{\mathfrak}
\newcommand{\wh}{\widehat}
\newcommand{\br}{\langle}
\newcommand{\kt}{\rangle}


\def\lsim{\mathrel{\mathop  {\hbox{\lower0.5ex\hbox{$\sim$}
\kern-0.8em\lower-0.7ex\hbox{$<$}}}}}
\def\gsim{\mathrel{\mathop  {\hbox{\lower0.5ex\hbox{$\sim$}
\kern-0.8em\lower-0.7ex\hbox{$>$}}}}}

\def\nn{\\  \nonumber}
\def\de{\partial}
\def\brf{{\mathbf f}}
\def\bbf{\bar{\bf f}}
\def\bF{{\bf F}}
\def\bbF{\bar{\bf F}}
\def\bA{{\mathbf A}}
\def\bB{{\mathbf B}}
\def\bG{{\mathbf G}}
\def\bI{{\mathbf I}}
\def\bM{{\mathbf M}}
\def\bY{{\mathbf Y}}
\def\bX{{\mathbf X}}
\def\bS{{\mathbf S}}
\def\bb{{\mathbf b}}
\def\bh{{\mathbf h}}
\def\bg{{\mathbf g}}
\def\bla{{\mathbf \la}}
\def\bmu{\mathbf m }
\def\by{{\mathbf y}}
\def\bmu{\mbox{\boldmath $\mu$} }
\def\bsig{\mbox{\boldmath $\sigma$} }
\def\bunity{{\mathbf 1}}
\def\cA{{\cal A}}
\def\cB{{\cal B}}
\def\cC{{\cal C}}
\def\cD{{\cal D}}
\def\cF{{\cal F}}
\def\cG{{\cal G}}
\def\cH{{\cal H}}
\def\cI{{\cal I}}
\def\cL{{\cal L}}
\def\cN{{\cal N}}
\def\cM{{\cal M}}
\def\cO{{\cal O}}
\def\cR{{\cal R}}
\def\cS{{\cal S}}
\def\cT{{\cal T}}
\def\eV{{\rm eV}}
%
\hfill \hbox{CERN-TH-2021-210}
\vskip 3.0cm

\large
 \begin{center}
 {\Large \bf  Angular momentum loss in gravitational scattering, \\ radiation reaction, 
 and the Bondi gauge ambiguity}

 \end{center}

 \vspace{0.1cm}

 \vspace{0.1cm}

  \begin{center}
{\large Gabriele Veneziano}\footnote{E-mail: \, gabriele.veneziano@cern.ch}
\\
{\it Theory Department, 
CERN, CH-1211 Geneva 23, Switzerland  \\ Coll\'ege de France, 11 place M. Berthelot, 75005 Paris, France}
\end{center}

  \begin{center}
{\large Gregory A. Vilkovisky}\footnote{E-mail: \, vilkov@lebedev.ru}
\\
{\it P.N. Lebedev Physical Institute of the Russian Academy of Sciences, 119333, \\  Leninski ave, 53, Moscow, Russia}
\end{center}

\vspace{1cm}
\begin{abstract}
\small

 Recently, Damour computed the radiation reaction on gravitational scattering as the (linear) response
to the angular momentum loss which he found to be of ${\cal O}(G^2)$ in the gravitational constant.
This is a puzzle because any amplitude calculation would predict both radiated energy and radiated angular momentum to start
only at ${\cal O}(G^3)$. Another puzzle is that the resultant radiation reaction, of ${\cal O}(G^3)$,
is nevertheless correct and confirmed by a number of direct calculations. We ascribe these puzzles to
the BMS ambiguity in defining angular momentum. The loss of angular momentum is to be counted out from
the  ADM value and, therefore, should be calculated in the so-called canonical gauge under the BMS
transformations in which the remote-past limit of the Bondi angular momentum coincides with the ADM
angular momentum. This calculation correctly gives the ${\cal O}(G^3)$ radiative loss. On the other hand, we introduce
a gauge in which the Bondi light cones tend asymptotically to those emanating from the center of mass
world line. We find that the angular momentum loss in this gauge is precisely the one used by Damour
for his radiation reaction result. We call this new gauge "intrinsic" and argue that, although the
radiated angular momentum is to be computed in the canonical gauge, any mechanical calculation
of gauge-dependent quantities -- such as angular momentum --  gives the result in the intrinsic gauge..
Therefore, it is this gauge that should be used in the linear response formula. This solves the puzzles
and establishes the correspondence between the intrinsic mechanical calculations and the Bondi formalism.

\end{abstract}

\newpage

\section{Introduction and summary}
\setcounter{equation}{0}

Two particles interact gravitationally and produce gravitational waves. This gravitational scattering problem is presently at the center of attention of many researchers, particularly in what concerns the problem of adding radiation reaction effects to the conservative potential dynamics. Such effects appear for the first time at the $3PM$ (or ${\cal O}(G^3)$) level and, in that context, have been found \cite{DiVecchia:2020ymx} to be essential for recovering the smooth ultra-relativistic limit first obtained in \cite{Amati:1990xe} and recently confirmed in \cite{DiVecchia:2020ymx} and \cite{Bern:2020gjj}. The calculation in \cite{DiVecchia:2020ymx} was carried out in the simpler framework of ${\cal N}=8$ massive supergravity and pointed to the need to include the full soft region in the loop integrals (the previously considered potential region \cite{Bern:2019nnu}, \cite{Bern:2019crd} being unable to include radiation reaction).
In an impressive paper \cite{Damour:2020tta} Damour found a smart shortcut for evaluating the ${\cal O}(G^3)$ radiation reaction (in the purely gravitational case). His paper is the subject of our considerations below.

The theory of gravitational radiation is the Bondi formalism introduced
in \cite{Bondi:1960jsa}, \cite{Bondi:1962px}, extended in \cite{Sachs:1962zza}, \cite{Sachs:1962wk}, and amplified by Penrose in \cite{Penrose:1964ge}.
A recent review can be found in \cite{Madler:2016xju}.
We also recommend \cite{Flanagan:2015pxa}, and especially  \cite{Bonga:2018gzr} where very clear and detailed equations
pertaining to the Bondi formalism are presented.

The radiation occurs at the future null
infinity ($\cal I^+$) which is the product
of the time-$u$ axis and the celestial 2-sphere $S$. The time $u$ labels the null hypersurfaces whose generators are the light rays that, when traced to the future,
come to the asymptotically flat infinity. $\cal I^+$ can be regarded as the future limit along these rays.

The two radiative degrees of freedom of the gravitational field can
be packaged into a symmetric traceless tensor $f_{ab}$, $a,b=1,2$, on the
2-sphere $S$, the {\it shear} tensor in the Bondi metric. Differentiated with respect to $u$, its two components
are the Bondi-Sachs news functions
describing  gravitational radiation. The fluxes of energy,
momentum, and angular momentum  due to radiation are given by the
expressions
 \be{Eloss}
  \partial_u M = - \frac{1}{32 \pi G} \int (\partial_u f_{ab}) (\partial_u f^{ab}) d^2 S \, ,
	\ee
   \be{Ploss}
  \partial_u P^i = - \frac{1}{32 \pi G} \int (\partial_u f_{ab}) (\partial_u f^{ab}) n^i d^2 S \, ,
  \ee
    \be{Jloss}
   \partial_u J^i = - \frac{3}{16 \pi G} \epsilon^i{}_{jk}  \int (n^{[j} D_a n^{k]}) \left(\frac16 (\partial_u f_{bc})(D^b f^{ac}) - \frac12 (\partial_u f^{ac})(D^b f_{bc})\right) d^2 S 
  \ee
where the integrals are over the (unit) 2-sphere $S$. The integration measure and
contractions are with respect to the standard metric on the unit 2-sphere (denoted below by $\Omega_{ab}$),
$D_a$ is the covariant derivative with respect to this metric,
and $n^i$ is the direction 3-vector living on the 2-sphere.

The news functions $\partial_u f_{ab}$ are ${\cal O}(G^2)$ since this is the lowest order at which the scattering effects manifest themselves in the metric. Hence the fluxes of energy and momentum are ${\cal O}(G^3)$. The statement persistently appears in the literature that the flux of angular momentum is by an order of $G$ larger:
\be{falseJ}
\partial_u J^i = {\cal O}(G^2)\, .
\ee
The reason is  that expression \eqref{Jloss}, as distinct from \eqref{Eloss} and \eqref{Ploss}, contains not only the news functions. It contains also $f_{ab}$ undifferentiated with respect to $u$. Damour \cite{Damour:2020tta} calculated the undifferentiated $f_{ab}$ and found
\be{falsef}
f_{ab}={\cal O}(G)\, .
\ee
Hence \eqref{falseJ}. The statement goes back to much earlier work by Damour and Deruelle \cite{Damour:1981bh}. There too the angular momentum loss is of lower order in the coupling constant than the energy loss. It got to the point where at a recent workshop there appeared a graviton having zero energy and robust angular momentum! This state of affairs is one of our concerns in the present paper.

It would not be difficult to correct the
statement above if it were not for another fact. Bini and 
Damour \cite{Bini:2012ji} derived a linear response formula for the scattering angle. The scattering angle $\chi$ is
divided into two contributions: $\chi^{\rm cons}$ which results
from the conservative dynamics and $\chi^{\rm rad}$ which is the
radiation-reaction contribution. The linear response formula
expresses $\chi^{\rm rad}$ through $\chi^{\rm cons}$ as follows:
\be{BiniDamour}
\chi^{\rm rad}(M,J)=-\frac{1}{2}\frac{\partial \chi^{\rm cons}}{\partial M} M^{\rm rad} -\frac{1}{2}\frac{\partial \chi^{\rm cons}}{\partial J} J^{\rm rad} 
\ee                                
where $M^{\rm rad}$ and $J^{\rm rad}$ are the total radiated energy
and angular momentum
\be{JMrad}
M^{\rm rad} =- \int\limits_{-\infty}^\infty du\,\partial_u M ~,~  J^{\rm rad}  = - \int\limits_{-\infty}^\infty du\,\partial_u J \, .
\ee
Damour \cite{Damour:2020tta} inserted 
into this formula the apparently incorrect ${\cal O}(G^2)$ flux of angular momentum and obtained the correct scattering angle! 
In subsequent work his result was re-derived by a very different shortcut \cite{DiVecchia:2021ndb} using soft-graviton theorems and analyticity arguments. It was finally confirmed by other, brute-force, calculations \cite{DiVecchia:2021bdo}, \cite{Herrmann:2021tct}, \cite{Bjerrum-Bohr:2021din}, \cite{Brandhuber:2021eyq}.

The purpose of the present work is to propose a resolution of the above puzzles.
Since Damour's shear \eqref{falsef} is ${\cal O}(G)$, we had to derive and
study the ${\cal O}(G)$ metric, i.e., the metric  
generated by a collection of non interacting particles at the lowest order in $G$. The relevant part of this study is presented
below, and here is the summary of its outcome. 
In its Bondi form, the ${\cal O}(G)$ metric contains an
arbitrariness which is none other than the BMS 
ambiguity \cite{Sachs:1962zza}, \cite{Madler:2016xju}  
which affects the shear tensor and the Bondi angular momentum\footnote {See \cite{Ashtekar:2019rpv} for another recently discussed implication of this ambiguity.}.
The solution of the puzzles is in the possible gauge choices 
with respect to the BMS transformations (specifically, to the supertranslations).

It is known (see e.g. \cite{Flanagan:2015pxa}) that the BMS supertranslations 
can be parametrized by the value of the shear tensor at $u=-\infty$, and
there is a ``canonical'' gauge in which the BMS ambiguity is fixed
by the condition
\be{canonical}
f_{ab}\bigl|_{u=-\infty} = 0. 
\ee                                       
In the ${\cal O}(G)$ metric, the shear tensor is $u$-independent, and we present
explicitly the supertranslation that turns it into zero.

The canonical gauge is of paramount importance because in this gauge, and only in this gauge, the Bondi angular momentum at $u=-\infty$ coincides with the ADM angular momentum. This statement has been proven in \cite{Ashtekar1} for stationary spacetimes. Soon after, the reasoning was generalized \cite{Ashtekar:1979xeo} to show that the past limit of the Bondi 4-momentum is the ADM 4-momentum for general radiating spacetimes. It is straightforward to combine the two arguments to show that the past limit of the Bondi angular momentum computed in the canonical gauge equals the ADM angular momentum  also in radiating spacetimes \cite{AA}. Because the latter reference is not yet available in print, we presently take the  above statement as an assumption. 
The following argument supports this assumption. There is no question that the loss of angular momentum is to be counted out from the initial ADM value.  There is also no doubt that gravitational radiation (seen as graviton emission) starts at ${\cal O}(G^3)$ in a purely gravitational collision~\footnote{This parallels the well-known fact that, in QED, photon emission in an $e^+ e^-$ collision starts at ${\cal O}(\alpha^3)$.}. Since the ${\cal O}(G)$ term of $f_{ab}$ sits only in its past limit, this means that the radiated angular momentum should be calculated in the canonical gauge. We see that the above statement/assumption reconciles classical GR reasoning with the $G$ power counting one derives from scattering amplitudes.

There is, however, another gauge relevant for the scattering problem
(and possibly elsewhere). It is fixed by the requirement that the Bondi
light cones coincide at $\cal I^+$ with the light cones emanating
from the world line of the center of mass of the particles' system.
We call this gauge ``intrinsic'' because it is attached to the dynamics
of particles. When a gauge-dependent quantity such as angular
momentum is calculated by working with the dynamical equations in the
center of mass frame, the result is obtained in the intrinsic gauge.
The Bini-Damour formula is derived by this kind of calculations.
Therefore, the angular momentum loss to be inserted in this formula
should be taken in the intrinsic gauge rather than in the canonical
gauge.

It turns out that the Damour's shear tensor \eqref{falsef} is precisely the one
obtained in the intrinsic gauge. This explains both his correct result for the
scattering angle and his incorrect  identification of the radiated angular momentum. The Bini-Damour formula is very valuable because
it is a rare case where the conservation laws in field theory
help to solve a dynamical problem. But one should know how to use
this formula.

The calculations in the paper \cite{Damour:1981bh} are also of the
intrinsic mechanical kind. Therefore, the result for the angular momentum loss 
is obtained in the intrinsic gauge. To obtain the radiated
angular momentum, it should be supertranslated to the canonical
gauge.

The flux of angular momentum in the intrinsic gauge gives the loss of mechanical $J$. It is larger than the radiated $J$. A large amount of it is transferred to the non-radiative part of the gravitational field. This is the so-called Schott term. In \eqref{Jloss} it has the form of a total time derivative which, unlike for elliptic motion, does not integrate to zero for the scattering case. Schott terms of this form arise already in the PN expansion (see e.g. \cite{Blanchet:2013haa}).

Thus the solution of the puzzle lies in the way the limit of  $\cal I^+$ is taken. To each point of  $\cal I^+$ there comes a two-parameter family of parallel light rays. At leading order they are indistinguishable, but shear and angular momentum are sensitive to which representative of the family is taken for the limit. This is the reason behind the BMS ambiguity. Damour's  result corresponds to the BMS gauge in which the relevant light rays emanate from the center-of-mass world line (to be more precise, tend to them asymptotically). This defines the intrinsic gauge.

The rest of the paper contains details of the above consideration.
In Section 2 we present the ${\cal O}(G)$ metric and calculate its Bondi shear. In the process of derivation we discover the arbitrariness in the Bondi metric. In Section 3 we provide an explanation of this arbitrariness. In Section 4, the intrinsic gauge is introduced, and both gauges, intrinsic and canonical, are discussed. In Section 5 we consider Damour's \cite{Damour:2020tta} shear and find that it is exactly the Bondi shear in the intrinsic gauge. The consequences of this fact were discussed above.

\section{The ${\cal O}(G)$ metric. Shear} 
\setcounter{equation}{0}

The ${\cal O}(G)$ metric obtained initially in the coordinate-independent form
\be{sol1}
 g^{\mu\nu} = g^{\mu\nu}_{{\rm flat}} + \delta g^{\mu\nu} ~,~  \delta g^{\mu\nu} = {\cal O}(G),
\ee
and next specialized to the Minkowski coordinates of $g^{\mu\nu}_{{\rm flat}}~$: $x^{\mu} = (x^0 , x^i)$ is of the form\footnote{This result can be obtained, up to a diffeomorphism, by directly solving the Einstein equations; it can also be obtained by judiciously performing a Lorentz boost of the Schwarzschild solution (see e.g. Appendix C of \cite{Bonga:2018gzr}).}
\be{Mmetric}
g^{\mu \nu} =\eta^{\mu\nu} - 4G \sum \frac{m}{\Gamma (x)} (v^{\mu}v^{\nu} + \frac12 \eta^{\mu\nu})   \, ,
\ee
\be{gamma}
\Gamma(x) = \sqrt{(x^{\mu} -c^{\mu})(x^{\nu} -c^{\nu})\Pi_{\mu \nu}} \, , 
\ee
\be{projection}
\Pi_{\mu \nu} = \eta_{\mu\nu} + v_{\mu} v_{\nu}~~,~~ v^{\mu} \Pi_{\mu \nu} = 0\, .
\ee
In \eqref{Mmetric}, and hereafter, the expression following the $\sum$ sign is the contribution of a single particle of mass $m$, and $\sum$ denotes the sum of contributions of all particles. Operations on the indices $\mu, \nu, \dots$ are performed with the Minkowski  metric $\eta_{\mu\nu} = {\rm diag~}(-1,1,1,1)$, and $v^{\mu}$, $c^{\nu}$ are constant 4-vectors appearing in the particle's law of motion with respect to its proper time $s$:
\be{freemotion}
 x^{\mu}(s) = c^{\mu} +v^{\mu} s  ~,~ v^2 =-1  \, , 
\ee
\be{energy}
mv^\mu= (E, p^i)\, .
\ee
Here $E$ and $p^i$ are the particle's energy and momentum. 

We stress that, besides the ${\cal O}(G)$ approximation, the metric above is exact in the sense that it does not involve any large- or small-distance approximation.

Using standard definitions related to the behavior of the metric at spatial infinity, it is straightforward, though somewhat tedious, to check that the metric (\ref{Mmetric}) reproduces the correct expressions for the ADM mass/energy, linear and angular momentum of the system of non interacting particles. We find:
\be{ADMP}
M_{\rm ADM} = \sum E~~,~~ P^i_{\rm ADM} = \sum p^i ~~,~~ J^i_{\rm ADM} = \epsilon^i{}_{j k}  \sum  x^j p^k = \epsilon^i{}_{j k} \sum  c^j p^k 
\ee
in  agreement with expectations.

Before proceeding we fix the Lorentz frame, thus far arbitrary,  to be the center of mass  (c.m.) frame in which the total momentum vanishes:
\be{com}
 \sum  p^i = 0 \, ,
 \ee
and $x^i=0$ is the world line of the center of mass of the particles' system.

We need to transform the metric above to the Bondi coordinates
\be{BScoor}
 {\bf u}, {\bf r}, {\bf\Phi}^a~,~ a= 1, 2 
 \ee
 defined by the conditions
\be{BScond}
(\nabla {\bf u})^2 =0~,~( \nabla{\bf u}, \nabla {\bf \Phi}^a) = 0 ~,~ \det (\nabla{\bf \Phi}^a, \nabla{\bf \Phi}^b) = {\bf r}^{-4}\left( \det{\bf \Omega}_{ab}\right)^{-1} \, .
\ee
Here ${\bf\Phi}^a$ take values on the 2-sphere and label the generators of the null surfaces ${\bf u}={\rm const.}$, 
\be{Omegadef}
{\bf \Omega}_{ab}={\bf \Omega}_{ab}({\bf\Phi})
\ee
is the standard metric on the unit 2-sphere, and
${\bf r}$ is the luminosity (area) distance. We are interested in the angular components of the Bondi metric, ${\bf g}_{ab}$, the inverse of
\be{cond}
{\bf g}^{ab}=(\nabla{\bf\Phi}^a,\nabla{\bf\Phi}^b) \, .
\ee
In the limit of ${\cal I}^+$, ${\bf g}_{ab}$ is of the form
\be{gabexp}
{\bf g}_{ab} = {\bf r}^2\left({\bf \Omega}_{ab} + \frac{1}{{\bf r}} {\bf f}_{ab} +{\cal O}(\frac{1}{{\bf r}^2} )\right)~,~ {\bf r}\to\infty 
\ee
where ${\bf f}_{ab}$ is the shear tensor. By virtue of the definition of ${\bf r}$ above, this tensor satisfies the trace-free condition
\be{trace}
{\bf\Omega}^{ab} {\bf f} _{ab} = 0 \, ,
\ee 
and then its two independent components are the two radiative degrees of freedom of the gravitational field.

We shall look for the solution of eqs. \eqref{BScond} in the form
\be{rel}
{\bf u} =  u + \delta u~,~ {\bf r} = r + \delta r~,~ {\bf\Phi}^a = \phi^a + \delta \phi^a  
\ee
where 
\be{flat}
u, r, \phi^a~,~ a= 1, 2 
\ee
are flat-space Bondi coordinates for which we choose the surfaces $u = {\rm const.}$ to be the future light cones (in flat metric) emanating from the c.m. world line: 
\be{Msph}
u=x^0-r~,~r=\sqrt{\sum (x^i)^2}~,~x^i = r n^i(\phi)\, .
\ee
Here $n^i(\phi)$ is the direction 3-vector which already figured in Sect. 1, and, 
for later convenience, we introduce also the 4-vector $ n^{\mu} =(1, n^i)=-\eta^{\mu\nu}\nabla_\nu u$. In what follows, the scalar product $( n \cdot v)$ denotes $\eta_{\mu\nu} n^\mu v^{\nu}$.

For the ${\cal O} (G)$ corrections in \eqref{rel} we obtain the equations
\be{diffcondu}
 \partial_r \delta u = - \sum \frac{2 G m}{\Gamma} (n \cdot v)^2 \; ,
\ee
\be{diffconda}
 \partial_r \delta \phi^a = 
 \frac{1}{r^2} \Omega^{ab} \partial_b \delta u  + \frac1r \sum \frac{4 G m}{\Gamma} (n \cdot v) \Omega^{ab} \partial_b (n \cdot v) \;,
 \ee
 \be{deltar}
 -2  \delta r =  r D_a \delta \phi^a - r \sum \frac{2 G m}{\Gamma} (\Omega^{ab}  \partial_a (n \cdot v) \partial_b (n \cdot v) + 1)  \, .
\ee
Since the equations are differential, their solution contains integration ``constants", i.e., arbitrary functions of $u$ and $\phi^a$. Solving them asymptotically as an expansion at ${\cal I}^+$, we find\footnote{The coefficient of $\log r$ in \eqref{foru} is initially obtained as $\sum 2 G m ( n \cdot v)$ but, owing to the c.m. condition \eqref{com}, it reduces to $-\sum 2 G E$ and thus becomes angle-independent. The validity of the entire theory rests on this crucial fact.}
\be{foru}
 \delta u\Bigl|_{{\cal I}^+} = - \left(\sum 2 G E\right) \log r   -  \beta (u, \phi^a) + {\cal O}(\frac1r) \, , 
\ee
\be{forphi}
 \delta \phi^a\Bigl|_{{\cal I}^+} = \Omega^{ab} \left( \gamma_b(u, \phi^a) + \frac{1}{r} \partial_b  \beta (u, \phi^a)  \right)+ {\cal O}(\frac{1}{r^2})\; , 
\ee
\be{forr}
 2 \delta r\Bigl|_{{\cal I}^+} = - r D^a \gamma_a(u, \phi^a) +  \left( \sum 2 G E - D^2  \beta(u, \phi^a)\right)  + {\cal O}(\frac{1}{r}) 
\ee
 where $\beta$ and  $\gamma_a$  are the above mentioned  integration ``constants". As explained below, $\beta$ corresponds to the BMS supertranslation arbitrariness, while $\gamma_a$ represent another residual arbitrariness of the Bondi coordinates: the freedom to make the transformation $\phi^a \rightarrow f^a(\phi, u)$.

 It is now straightforward to calculate the Bondi metric as an asymptotic expansion near ${\cal I}^+$ with the presently needed accuracy. The arbitrary functions are restricted by the requirement that the Bondi metric have the correct flat-space limit at ${\cal I}^+$. This leads to a set of entangled equations for $\beta$ and $ \gamma_a$.
 They fix $ \gamma_a$ up to several constants, and, without loss of generality, we can set $\gamma_a = 0$.
 Then the solution for $ \beta$ is of the form
 \be{betacon}
 \beta = \beta_1(u) + \beta_2(\phi^a)
 \ee
 with arbitrary $\beta_1$ and $\beta_2$. The term $ \beta_1(u)$ can be absorbed by a trivial redefinition $ u \to f(u)$. The remaining function $ \beta_2(\phi^a)$ is the genuine supertranslation parameter.
 
 For the shear tensor we obtain the following result:
 \beqn{shear}
{\bf f}_{ab} =& \! - \!& \sum \frac{\displaystyle 4 G m}{\displaystyle (n \cdot v)} \left[D_a (n \cdot v) D_b ( n \cdot v) - \frac12 \Omega_{ab}D_c (n \cdot v) D^c ( n \cdot v) \right]  \nonumber \\
&\! + \!& (\Omega_{ab} D^2 - 2 D_a D_b) \beta   \, .
\eeqn
There is an order-$G$ term in the shear (recall \eqref{falsef}) but there is also a term containing an arbitrary function. The shear is gauge-dependent. We shall see below the implications of this fact.

 \section{Nature of the supertranslation ambiguity}
\setcounter{equation}{0}

We shall explain the arbitrariness in the Bondi metric by showing that it describes an arbitrary deflection of the central line ${\bf r}=0$ of the Bondi frame from the c.m. world line $r=0$ of the particles' system. For that purpose we need the exact solution of eq. \eqref{diffcondu} normalized at $r=0$. The function $\Gamma^2$ is a quadratic polynomial in $r$. Introducing a notation for its coefficients, we have
\be{Gammapar}
\Gamma^2 = r^2 (n \cdot v)^2 - 2 r a (n \cdot v) + z \;.
\ee
Then the solution reads
\be{deltau}
 \delta u = \omega(u, \phi^a) + \sum 2 G m  (n \cdot v)  \log\left[ \frac{a + \Gamma -r  (n \cdot v)}{a + \sqrt{z}} \right] 
\ee
where $\omega(u, \phi^a)$ is the integration ``constant"  defined as the solution at $r=0$. Expanding \eqref{deltau} at ${\cal I}^+$, we find the relation between $\omega$ and $\beta$ (the renormalization):
\be{betaconstr}
 \beta = - \omega - \sum 2 G m (n \cdot v) \log\left[ \frac{-2 (n \cdot v)}{a + \sqrt{z}} \right] \, .
\ee
Here $z$ depends only on $u$, and $a$ on both $u$ and angles. Their explicit form will not be needed here.

We need to find $\delta r$ in \eqref{rel} in the $r = 0$ limit. This amounts to solving for the corrections $\delta u$, $\delta \phi^a$, $\delta r$ anew starting from \eqref{deltau} and expanding this time at $r\to 0$ rather than at $r\to\infty$. The final result is simple and eloquent:
\be{rB}
{\bf r}\bigl |_{r=0} = \frac12 D^2 \omega\, .
\ee
 Hence the deviation of the line ${\bf r} = 0$ from the line $r=0$ is directly related to the supertranslation parameter $\omega$.
 
 Can we set this deviation to zero? This would require $\omega$ to be a function of $u$ only. But with $\omega= \omega(u)$ the relation \eqref{betaconstr} will force $\beta$ to be a complicated function of $u$ and the angles which is clearly incompatible with \eqref{betacon}.
 We conclude that it is impossible to make the world lines ${\bf r} = 0$ and $r=0$ coincide and attribute this impossibility to the fact that the c.m. world line $r=0$ is a geodesic in flat metric but not in the ${\cal O}(G)$ metric while the central line of the Bondi frame is a geodesic in the exact metric. In the next section we shall discuss to what extent the Bondi frame can be based on the c.m. world line.

 \section{Canonical and ``intrinsic" Bondi gauges}
\setcounter{equation}{0}

The right-hand side of \eqref{diffcondu} is, up to a factor of 2, $(\nabla u)^2$ in the ${\cal O}(G)$ metric. It is not vanishing but it is ${\cal O}(1/r)$ at ${\cal I}^+$. Therefore,
\be{utilde}
(\nabla u)^2 \Bigl |_{{\cal I}^+}=0
\ee
even in the ${\cal O}(G)$ metric. In other words, although in the compact domain the surfaces $u ={\rm const.}$ are not null, in the limit of  ${\cal I}^+$ they become null and remain the light cones emanating from the c.m. world line.

Consider now imposing the condition
\be{uUcond}
\nabla \delta u\,\Bigl |_{{\cal I}^+} = 0~,~{\rm i.e.},~  \nabla{\bf u} = \nabla u~\,{\rm at}~{\cal I}^+ \, .
\ee
As seen from eq. \eqref{foru}, it amounts to
\be{nablabeta}
\nabla\beta = 0 
\ee
and thus fixes the supertranslation arbitrariness up to $\beta={\rm const.}$.
Since the vector $\nabla{\bf u}$ is the null tangent to the Bondi light cones, this gauge condition requires that the Bondi light cones tend asymptotically to those emanating from the c.m. world line\footnote{Since the term with $\log r$ in \eqref{foru} is angle-independent, it does not affect the shape of the light cones. At a given $r$, it is a large additive constant. }.
We have seen that it is not possible to impose such a condition everywhere in spacetime but it can be imposed at ${\cal I}^+$, and this is sufficient to fix the gauge.

We call this gauge ``intrinsic" because it is attached to the motion of particles in their c.m. frame. The ADM or Bondi definition of the center of mass is far removed from the events in the interior of spacetime where the particles interact. Therefore, it fixes only the freedom of performing the Lorentz boosts at infinity but not the freedom of the BMS supertranslations. The intrinsic mechanical definition which uses the notion of the c.m. world line is stronger. Basing the Bondi frame on this world line would fix also the supertranslation arbitrariness. And it suffices to do so near ${\cal I}^+$.

This is the meaning of the intrinsic gauge. When a gauge-dependent quantity, such as angular momentum, is calculated by working with the dynamical equations in the c.m. frame, the result is obtained in the intrinsic gauge. Since the linear response formula \cite{Bini:2012ji} is derived by this kind of calculations, the angular momentum loss to be inserted into this formula should be taken in the intrinsic --rather than in the canonical-- gauge. 

It follows from eq. \eqref{shear} that in the intrinsic gauge 
\be{intrbeta}
\beta = 0\, ,
\ee
\be{intrf}
{\bf f}_{ab} = - \sum \frac{4 G m}{(n \cdot v)} \left[D_a (n \cdot v) D_b ( n \cdot v) - \frac12 \Omega_{ab}D_c (n \cdot v) D^c ( n \cdot v) \right] \, .
\ee
In the canonical gauge conversely
\be{canf}
{\bf f}_{ab}=0\, ,
\ee
\be{canbeta}
\beta = - \sum 2Gm (n\cdot v)\log (-(n\cdot v))\, .
\ee
One can check by a direct calculation that this $\beta$ represents precisely the supertranslation that turns the shear \eqref{shear} into zero.

\section{Damour's shear \cite{Damour:2020tta} }
\setcounter{equation}{0}

Consider the ${\cal O}(G)$ term of the metric in Minkowski coordinates, its spatial part
\be{spatial}
\delta g_{ik} = \sum\frac{4Gm}{\Gamma}\left(\frac{p_ip_k}{m^2}+\frac{1}{2}\delta_{ik}\right) 
\ee
and convert it into a 2D tensor 
\be{naiveF}
F_{ab}=(D_a n^i)(D_b n^k)\delta g_{ik}\, .
\ee
To see what this corresponds to in the Bondi coordinates, transform the metric from Minkowski to {\it flat-space} Bondi coordinates (marked below as ``Bondi"). For the angle components of the ``Bondi" metric one obtains
\be{angle "Bondi"}
{\rm ``Bondi"}~\, g_{ab}=(D_a x^i)(D_b x^k)(\delta_{ik} +\delta g_{ik})= r^2(\Omega_{ab}+F_{ab})
\ee
and notices that if one writes
\be{smallf}
F_{ab}=\frac{1}{r} f_{ab}~,~r\to\infty\, ,
\ee
then $f_{ab}$ plays the role of shear in the ``Bondi" metric. One might try to take it for one's shear but, because the ``Bondi'' $r$ is not correctly defined, the ``shear'' does not satisfy the trace-free condition:
\be{netrace}
\Omega^{ab}F_{ab}=(\delta^{ik}-n^in^k)\delta g_{ik}\ne 0\, .
\ee
However, one can correct $F_{ab}$:
\be{corF}
F_{ab}^{~\rm cor}=(D_a n^i)(D_b n^k)\delta g_{ik}^{\rm TT}~\, ,~\;  F_{ab}^{~\rm cor}=\frac{1}{r} f_{ab}^{~\rm cor} 
\ee
where $\delta g_{ik}^{\rm TT}$ is the transverse and traceless part of $\delta g_{ik}$. Since
\be{TT}
n^i\delta g_{ik}^{\rm TT}=0 ~\, ,~\;  \delta^{ik} \delta g_{ik}^{\rm TT}=0\, ,
\ee
$F_{ab}^{~\rm cor}$ satisfies the trace-free condition. The $f_{ab}^{~\rm cor}$ is Damour's \cite{Damour:2020tta} shear.

It is not difficult to calculate $f_{ab}^{~\rm cor}$ in \eqref{corF}. The result is expression \eqref{intrf}. Damour's  shear \cite{Damour:2020tta} is exactly the Bondi shear in the intrinsic gauge! This explains both the success of ref. \cite{Damour:2020tta} in the calculation of radiation reaction through the linear response formula and its failure  in obtaining the true angular momentum loss.

Let us finally comment on the calculation of \cite{Damour:1981bh} for which most of the remarks made about \cite{Damour:2020tta} remain true. We can add that, in the non-relativistic limit considered in \cite{Damour:1981bh}, the news functions (converted into a 3D tensor) can be obtained as the third time derivative of the quadrupole moment of the system\footnote{This relation remains valid in a fully relativistic version \cite{Mirzabekian:1998ha}.}:
\be{quadrupole}
\partial_u f_{ij} = \frac{2G}{3} \partial^3_{uuu} Q_{ij}^{\rm TT}\, .
\ee
The solution of this equation for shear is
\be{quadrupoleshear}
f_{ij} = \frac{2G}{3} \partial^2_{uu} Q_{ij}^{\rm TT}+ c_{ij} (\phi)\, ,
\ee
and the integration ``const.'' $c_{ij}(\phi)$ cannot be ignored. For the purpose of obtaining the true angular momentum flux it should be fixed by the requirement $f_{ij}(u \to - \infty) =0$ of the canonical gauge.

{\large \bf Acknowledgments}

The authors are grateful to Abhay Ashtekar for his invaluable help throughout this work, including  private communications on Refs. \cite{Ashtekar1}, \cite{Ashtekar:1979xeo},  \cite{AA}, to Eric Poisson for instructive communications, and to Thibault Damour for clarifying critical remarks. One of us (Gabriele Veneziano) would like to acknowledge useful (albeit remote) discussions with several participants at the  workshop ``Gravitational scattering, inspiral, and radiation" held at the Galileo Galilei Institute in April-May 2021. We also acknowledge useful remarks from two anonymous reviewers that resulted in improvements of the original text.

\vspace{4mm}

{\bf Notes Added }
\begin{itemize}
\item After completion of this paper we were informed by R. Oliveri of the existence of  a paper \cite{Blanchet:2020ngx} where a similar coordinate transformation from harmonic to Bondi coordinates is performed in the context of the multipolar post-Minkowskian approximation. Unfortunately, the issue of the Bondi gauge dependence of the radiated angular momentum is not addressed in that paper.
\item
 Two recent  \cite{Manohar:2022dea}, \cite{DiVecchia:2022owy} amplitude-based $\it{derivations}$ of Damour's result have appeared, in which the ${\cal O}(G^2)$ contribution of exactly zero-energy gravitons (whatever that means) is included. Our interpretation of such ``gravitons" (as  corresponding to  a  non-radiative component of the gravitational field) is in line with that of the authors.
 \end{itemize}

\end{document}